\documentclass[aps,prd,preprint,groupedaddress]{revtex4}
\usepackage{graphicx}

\begin{document}

\preprint{OU-HEP-518}

\title{The generalized parton distribution functions and the nucleon
spin sum rules in the chiral quark soliton model}


\author{M.~Wakamatsu and H.~Tsujimoto}
\email[]{wakamatu@phys.sci.osaka-u.ac.jp} \ 
\email[]{tujimoto@kern.phys.sci.osaka-u.ac.jp}
\affiliation{Department of Physics, Faculty of Science, \\
Osaka University, \\
Toyonaka, Osaka 560-0043, JAPAN}



\begin{abstract}
The theoretical predictions are given for the forward limit of the
unpolarized spin-flip isovector generalized parton distribution function
$(E^u - E^d)(x, \xi, t)$ within the framework of the chiral quark
soliton model, with full inclusion of the polarization of Dirac sea
quarks. We observe that $[(H^u - H^d) + (E^u - E^d)](x,0,0)$
has a sharp peak around $x=0$, which we interpret as a signal
of the importance of the pionic $q \bar{q}$ excitation with large
spatial extension in the transverse direction.
Another interesting indication given by the predicted distribution
in combination with Ji's angular momentum sum rule is that the
$\bar{d}$-quark carries more angular momentum than the $\bar{u}$-quark
in the proton, which may have some relation with the physics of the
violation of the Gottfried sum rule.
\end{abstract}

\pacs{12.39.Fe, 12.39.Ki, 12.38.Lg, 13.40.Em}

\maketitle


\section{Introduction}

 A distinguishable feature of the chiral quark soliton model (CQSM) as
compared with many other effective model of baryons, like the naive
quark model or the MIT bag model (at least in its most primitive version),
is that it is a field theoretical model which takes account not only
of three valence quarks but also of infinitely many Dirac sea-quark
degrees of freedom \cite{DPP88},\cite{WY91}.
As emphasized by Diakonov in his recent review \cite{D04},
this feature is essential in explaining the so-called
"nucleon spin crisis" \cite{EMC88},\cite{EMC89} as well as the
quite large experimental value of the $\pi N$ sigma
term \cite{O00},\cite{PSWA02} : what he calls the two stumbling
blocks of the naive quark models.
In fact, the nucleon spin sum rule within the CQSM was first 
derived in \cite{WY91}.
It was shown there that the sizable amount of the
nucleon spin comes from the orbital angular momentum carried by Dirac
sea quarks. The way how the CQSM can explain the huge experimental
value of $\pi$N sigma term is also very interesting.
It predicts that only a small portion 
of the large $\pi$N sigma term is due to the main constituents of the 
nucleon, i.e. the three valence quarks, and the dominant contribution 
originates from the Dirac sea quarks \cite{DPP89},\cite{W92}.
Moreover, it was recently found that the Dirac-sea contribution
to the $\pi$N sigma term resides in a peculiar delta-function type 
singularity at $x = 0$ in the chiral-odd twist-3 distribution function 
$e(x)$ of the nucleon \cite{ES03}\nocite{WO03}\nocite{S03}--\cite{OW04}.
Also clarified there is that this delta-function singularity of
$e(x)$ is a rare manifestation of the nontrivial vacuum structure
of QCD, characterized by the non-zero quark condensate, in a
baryon observable. The superiority of the CQSM manifests even more
drastically in high-energy observables.
It is almost only one effective model that
can give reliable predictions for the quark and antiquark distribution
functions of the nucleon satisfying the fundamental field theoretical
restrictions like the positivity of the antiquark distribution
functions \cite{DPPPW96}\nocite{DPPPW97}\nocite{WGR96}\nocite{WGR97}
\nocite{WK98}\nocite{WK99}--\cite{PPGWW99}.

Coming back to the nucleon spin problem, we claim that the CQSM already 
gives one possible solution to it. The physical reason why this model 
predicts small quark spin fraction, or large orbital angular 
momentum is clear. It is connected with the basic nucleon picture of this 
model, i.e. ``rotating hedgehog''. Naturally, this unique nucleon picture 
takes over that of the Skyrme model. Immediately after the EMC
measurement, Ellis, Karliner and Brodsky showed that this unique model
predicts $\Delta \Sigma = 0$, i.e. vanishing quark spin
fraction \cite{BEK88}.
An important difference between these two intimately connected
models (the Skyrme model as an effective pion theory and the CQSM
as an effective quark model) should not be overlooked, however.
In the CQSM, $\Delta \Sigma$ receives small but definitely nonzero
contribution from the three valence quarks forming
the core of the nucleon. (Another way to get $\Delta \Sigma \neq 0$
is the generalization of the Skyrme model so as to include the
short range fields like the vector mesons \cite{JPSSW90}.)

At any rate, an interesting solution to the nucleon spin puzzle,
provided by the chiral soliton picture of the nucleon, emphasizes the
importance of the orbital angular momentum of quark and
antiquarks \cite{WY91}.
On the other hand, there is another completely different scenario
giving a possible solution to the nucleon spin puzzle.
It claims that the small quark spin fraction 
is compensated by the large gluon polarization (or the gluon 
orbital angular momentum.) Which scenario is favored by nature is
still an unsolved question, which must be answered by some experiments 
in the future. An experimental test of the first scenario has been
thought to be an extremely difficult task, because the quark orbital
angular momentum in the nucleon was not believed to be an experimentally
observable quantity.
The situation has changed drastically after Ji's
proposal \cite{Ji97A}\nocite{Ji97B}\nocite{HJL98}--\cite{Ji98}.
He showed that the quark total angular momentum and also
the quark orbital angular 
momentum in the nucleon can in principle be extracted through the 
measurement of the deeply virtual Compton scattering (DVCS) cross
sections in combination with analyses of the standard inclusive
reactions. 
The key quantity here is the unpolarized spin-flip generalized parton 
distribution function $E^q (x, \xi, t)$ appearing in the DVCS cross 
section formula. Especially interesting in the context of the above
argument is its forward limit $E^q (x,0,0) \equiv 
\lim_{t \rightarrow 0, \xi \rightarrow 0} E^q (x, \xi, t)$.
It satisfies the following second moment 
sum rule :
\begin{equation}
 \frac{1}{2} \,\int_{-1}^1 \,x [ f_1^q(x) + E^q (x, 0, 0) ] \,d x = J^q ,
\end{equation}
which is widely known as Ji's angular momentum sum
rule \cite{Ji97A}-\cite{Ji98}.
Here $f_1^q(x)$ is the standard unpolarized quark (and antiquark)
distribution  function of flavor $q$, while $J^q$ is the total angular
momentum of quarks (and antiquark) with flavor $q$.
Since $f_1^q(x)$ is already well known, the new experimental knowledge
of $E^q (x,0,0)$ would completely determine $J^q$.
Combining it with the available knowledge of the longitudinal quark
polarization $\Delta \Sigma$, together with the relation
\begin{equation}
 J^q = L^q + \frac{1}{2} \Delta \Sigma ,
\end{equation}
this opens up the possibility to extract the quark orbital angular
momentum in the nucleon purely experimentally.
Very recently, Ossmann et al. reported a very interesting calculation of 
$E^q (x, 0, 0)$, or more precisely the isoscalar combination
$E^u (x,0,0) + E^d (x,0,0)$, within the framework of the
CQSM \cite{OPSUG04}. They found that the
contribution of Dirac-sea quarks to $E^u (x,0,0) + E^d (x,0,0)$
dominates over that of the three valence quarks in the 
small $x$ region. Especially interesting is their finding that the
Dirac-sea contribution to $E^u (x,0,0) + E^d (x,0,0)$ has an $1/x$
singularity around $x = 0$ in the chiral limit.
Because of this peculiar feature, it turns out that the Dirac-sea term
gives negligible contribution to the first moment, 
i.e. $\int_{-1}^1 [ E^u (x,0,0) + E^d (x,0,0) ] d x$, or equivalently
to the isoscalar anomalous magnetic moment sum rule,
while it gives a sizable
contribution to the second moment sum rule, i.e. 
$\int_{-1}^1 \,x \,[ E^u (x,0,0) + E^d (x,0,0) ] \,d x$,
or to the quark angular momentum sum rule.
They also investigated the second moment sum rule for 
$E^u (x,0,0) + E^d (x,0,0)$ within the CQSM, and confirmed that it
reduces to the nucleon spin sum rule first derived in \cite{WY91}.
This was an expected result, since, in the CQSM, any physical
observables, including the nucleon spin, is saturated by the quark
field alone, (the quark intrinsic spin and the quark 
orbital angular momentum in the present case), and since it is
already known that the model satisfies the energy momentum sum rule
as well \cite{DPPPW97}.
Although only the flavor singlet combination appears in the total nucleon 
spin sum rule, we also need the independent isovector combination 
$E^u (x,0,0) - E^d (x,0,0)$, to make a flavor decomposition of
the quark angular momentum.
The isovector combination $E^u (x,0,0) - E^d (x,0,0)$ has already been
addressed partially, but in an incomplete
way \cite{PPPBGW98}\nocite{GPV01}--\cite{SBR02}.
The purpose of the present study is 
to carry out more complete investigation of this quantity.
We shall show the results of exact numerical calculation of this quantity 
without recourse to the derivative expansion type
approximation \cite{PPPBGW98},\cite{GPV01}. We also 
investigate the first and the second moment sum rule of
$E^u (x,0,0) - E^d (x,0,0)$, which is expected to give valuable
information on the theoretical consistency of 
the model.  

The paper is organized as follows. In sect.II, we briefly summarize
some basic properties of the unpolarized generalized parton
distribution functions (GPDF) necessary for our later discussion.
Sect.III is devoted to the theoretical analyses of the unpolarized
GPDF based on the CQSM. The main concern here is the first and
the second moments of the isovector spin-flip unpolarized
GPDF $E^{(I=1)}(x,\xi,t)$, which we know has intimate connection with the
nucleon isovector magnetic moment and the isovector combination of
the quark spin fraction of the nucleon.
Next, in sect.IV, we shall present the CQSM predictions for the
forward limit $E^{(I=1)}(x,0,0)$ of the isovector spin-flip
unpolarized GPDF. We also show the detailed numerical contents
of the first and the second moment sum rules of this quantity.
Finally, we summarize our findings in sect.V.

\section{General properties of the unpolarized GPDF}

 Here, we briefly summarize some important features of the
unpolarized quark generalized parton distribution functions
$H^q (x, \xi, t)$ and $E^q (x, \xi, t)$ with flavor $q$,
which are necessary for later discussion. They are defined by
\begin{eqnarray}
 \int \frac{d \lambda}{2 \pi} \,e^{i \lambda x} \,
 \langle \mbox{\boldmath $P$}^{\prime}, s^{\prime} | \bar{\psi}_q 
 \left(-\frac{\lambda n}{2} \right) 
 \not\!n \psi_q \left(\frac{\lambda n}{2}\right) | 
 \mbox{\boldmath $P$}, s \rangle
 &=& H^q (x, \xi, t) \bar{U} (\mbox{\boldmath $P$}^{\prime}, s^{\prime})
 \not\!n 
 U (\mbox{\boldmath $P$}, s) \nonumber \\
 &+& E^q (x, \xi, t) \bar{U} (\mbox{\boldmath $P$}^{\prime}, s^{\prime}) 
 \frac{i \sigma^{\mu \nu} n_{\mu} \Delta_{\nu}}{2 M_N}
 U (\mbox{\boldmath $P$},s) . \ \ \ 
\end{eqnarray}
Here (and hereafter) we omit the light-cone gauge link, for brevity. 
We use the standard notation :
\begin{equation}
 \Delta = P^{\prime} - P, \ 
 t = \Delta^2, \ 
 \xi = -\frac{1}{2} n \cdot \Delta ,
\end{equation}
with $n$ the light-like vector satisfying the relations.
\begin{equation}
 n^2 = 0, \ 
 n \cdot (P^{\prime} + P) = 2 ,
\end{equation}
It is a well known fact that the first moments of $H^q (x, \xi, t)$ and
$E^q (x, \xi, t)$ reduce to the Dirac and Pauli form factors,
respectively
\begin{eqnarray}
 \int_{-1}^1 H^q (x, \xi, t) \,d x &=& F_1^q (t), \\
 \int_{-1}^1 E^q (x, \xi, t) \,d x &=& F_2^q (t) .
\end{eqnarray}
For convenience, we introduce the isoscalar and isovector combinations 
as follows :
\begin{eqnarray}
 H^{(I = 0)} (x, \xi, t) &\equiv& H^u (x, \xi, t) + H^d (x, \xi, t) ,\\
 H^{(I = 1)} (x, \xi, t) &=& H^u (x, \xi, t) - H^d (x, \xi, t) ,
\end{eqnarray}
and similarly for $E^{(I=0)}(x,\xi,t)$ and $E^{(I=1)}(x,\xi,t)$.
(In the present paper, we neglect the strange quark degrees of freedom,
and confine to the two flavor case.)
The forward limit $(\xi \rightarrow 0, t \rightarrow 0)$ of the first 
moment sum rule then give
\begin{eqnarray}
 &\,& \int_{-1}^1 H^{(I = 0)} (x, 0, 0) \,d x = 3 , \\
 &\,& \int_{-1}^1 H^{(I = 1)} (x, 0, 0) \,d x = 1 , \\
 &\,& \int_{-1}^1 E^{(I = 0)} (x, 0, 0) \,d x = \kappa^u + \kappa^d 
 = 3 (\kappa^p + \kappa^n) = 3 \kappa^{(I = 0)} , \\
 &\,& \int_{-1}^1 E^{(I = 1)} (x, 0, 0) \,d x = \kappa^u - \kappa^d 
 = \kappa^p - \kappa^n = \kappa^{(I = 1)} .
\end{eqnarray}
Here $\kappa^u$ and $\kappa^d$ stand for the anomalous magnetic moments
of the $u$- and $d$-quarks, while $\kappa^p$ and $\kappa^n$ are those
of the proton and neutron.
The first moment sum rule of $E(x, 0, 0)$ has especially interesting
physical interpretation. Namely, $E(x, 0, 0)$ gives the distribution
of the nucleon anomalous magnetic moments in the Feynman
momentum $x$ space not in the ordinary coordinate space.
Also noteworthy is the second moment sum rule given as
\begin{equation}
 \frac{1}{2} \,\int_{-1}^1 x (H^{(I = 0)} + E^{(I = 0)}) (x, 0, 0) d x 
 = J^{(I = 0)} = J^u + J^d  , \label{angsum0}
\end{equation}
which is known as Ji's quark angular momentum sum
rule \cite{Ji97A}-\cite{Ji98}. Here, 
$J^u + J^d$ represents the total quark (spin and orbital
angular momentum) contribution to the nucleon spin. 
The forward limit of $H^q(x, \xi, t)$ is known
to reduce to the standard unpolarized distribution function,
which is rather precisely known by now. On the other hand,
the forward limit of $E^q(x, \xi, t)$ is believed to be extracted
from the analysis of the so-called deeply virtual Compton scatterings
on the nucleon target \cite{Ji97A}-\cite{Ji98}.
This means that the total quark angular momentum fraction of the nucleon 
spin can be determined purely experimentally. Subtracting the 
known value of the quark intrinsic spin fraction
$\Delta \Sigma^{(I=0)}$ of the nucleon. we can thus know the
quark orbital angular momentum fraction of the total nucleon spin as well.
Furthermore, making a different flavor combination 
(isovector combination) from (\ref{angsum0}),
one expects another sum rule :
\begin{equation}
 \frac{1}{2} \,\int_{-1}^1 x (H^{(I = 1)} + E^{(I = 1)}) (x, 0, 0) \,d x 
 = J^{(I = 1)} = J^u - J^d . \label{angsum1}
\end{equation}
Thus, with combined use of the isoscalar and isovector
sum rule, one would make a complete flavor decomposition of the
quark total angular momentum.

\section{unpolarized GPDF in the CQSM}

The theoretical expressions of $H (x, \xi, t)$ and $E (x, \xi, t)$ in the 
CQSM were already given in several previous
papers \cite{PPPBGW98}\nocite{GPV01}--\cite{SBR02},\cite{OPSUG04},
so that we do not repeat the detailed derivation here.
We describe only some main features and differences
for the sake of later discussion.
Here we closely follow the notation in \cite{OPSUG04},
and introduce the quantities
\begin{eqnarray}
 {\cal M}^{(I = 0)}_{s^{\prime} s} &\equiv& \int \frac{d \lambda}{2 \pi}
 e^{i \lambda x} \langle \mbox{\boldmath $P$}^{\prime}. 
 s^{\prime} | \bar{\psi}
 \left(-\frac{\lambda n}{2}\right) \not\!n 
 \psi \left(\frac{\lambda n}{2}\right) | 
 \mbox{\boldmath $P$}, s \rangle , \\
 {\cal M}^{(I = 1)}_{s^{\prime} s} &\equiv& \int \frac{d \lambda}{2 \pi}
 e^{i \lambda x} \langle \mbox{\boldmath $P$}^{\prime}, 
 s^{\prime} | \bar{\psi}
 \left(-\frac{\lambda n}{2}\right) \tau_3 \not\!n 
 \psi \left(\frac{\lambda n}{2}\right) | 
 \mbox{\boldmath $P$}, s \rangle .
\end{eqnarray}
The relations between these quantities and the generalized parton
distribution functions $H (x, \xi, t)$ and 
$E (x, \xi, t)$ are obtained most conveniently in the Breit frame.
They are given by 
\begin{eqnarray}
 {\cal M}^{(I = 0)}_{s^{\prime} s} &=& 2 \delta_{s^{\prime} s} 
 H^{(I = 0)}_E (x, \xi, t) - \frac{i \epsilon^{3 k l} \Delta^k}{M_N}
 (\sigma^l)_{s^{\prime} s} E^{(I = 0)}_M (x, \xi, t) , \label{hecomb} \\
 {\cal M}^{(I = 1)}_{s^{\prime} s} &=& 2 \delta_{s^{\prime} s} 
 H^{(I = 1)}_E (x, \xi, t) - \frac{i \epsilon^{3 k l} \Delta^k}{M_N}
 (\sigma^l)_{s^{\prime} s} E^{(I = 1)}_M (x, \xi, t) . \label{emcomb}
\end{eqnarray}
where
\begin{eqnarray}
 H^{(I=0/1)}_E (x, \xi, t) &\equiv& 
 H^{(I=0/1)} (x, \xi, t) + \frac{t}{4 M_N^2}
 E^{(I=0/1)} (x, \xi, t) ,\\
 E^{(I=0/1)}_M (x, \xi, t) &\equiv& 
 H^{(I=0/1)} (x, \xi, t) + E^{(I=0/1)} (x, \xi, t) . 
\end{eqnarray}
These two independent combinations of 
$H(x, \xi, t)$ and $E(x, \xi, t)$ can be extracted through the spin
projection of ${\cal M}^{(I)}$ as \cite{PPPBGW98},\cite{OPSUG04}
\begin{eqnarray}
 H^{(I)}_E (x, \xi, t) &=& \frac{1}{4} \mbox{tr} 
 \{ {\cal M}^{(I)} \} , \label{hetrace} \\
 E^{(I)}_M (x, \xi, t) &=& \frac{i M_N \epsilon^{3 b m} \Delta^b}
 {2 \mbox{\boldmath $\Delta$}_{\bot}^2} \mbox{tr} 
 \{ \sigma^m {\cal M}^{(I)} \} , \label{emtrace}
\end{eqnarray}
where ``$\mbox{tr}$'' denotes the trace over spin indices, while
$\mbox{\boldmath $\Delta$}_{\bot}^2 = \mbox{\boldmath 
$\Delta$}^2 - (\Delta^3)^2 = -t - (-2 M_N \xi)^2$.
Now, the r.h.s. of (\ref{hetrace}) and (\ref{emtrace})
can be evaluated within the framework of the CQSM.
Here, we briefly describe the basic features of
the CQSM leading to the theoretical expressions given below.
The CQSM is a relativistic mean field theory with hedgehog assumption
which breaks the rotation symmetry in addition to the translational
symmetry at the mean field level. Two zero-energy modes must be taken
into account to recover these symmetries.
To recover the translational invariance, we use an approximate method
which projects on the nucleon state with given center-of-mass momentum
$\mbox{\boldmath $P$}$ by integrating out the shift
$\mbox{\boldmath $x$}$ of the soliton center-mass
coordinate \cite{DPPPW96},\cite{DPPPW97} :
\begin{equation}
 \langle \mbox{\boldmath $P$}^{\prime} | \cdots | 
 \mbox{\boldmath $P$} \rangle 
 = \int d^3 \,\mbox{\boldmath $x$} \,
 e^{i (\mbox{\boldmath $P$}^{\prime} - \mbox{\boldmath $P$}) \cdot 
 \mbox{\boldmath $x$}} \cdots .
\end{equation}
Naturally, this procedure is justified only when the soliton is heavy 
enough and its center-of-mass motion is nonrelativistic. Another 
zero-energy mode corresponds to the soliton rotational motion.
As usual, the velocity of this time-dependent rotation is assumed to be
much slower than that of the intrinsic quark motions in the mean
field \cite{DPP88},\cite{WY91}.
This allows us to evaluate any nucleon observables in a
perturbation theory with respect to the soliton rotational
velocity $\Omega$.
This then leads to the following general structure of the theoretical 
expressions for nucleon observables in the CQSM. The leading contribution 
just corresponds to the mean field prediction, which is independent of 
$\Omega$. The next-to-leading order term takes account of the linear 
response of the intrinsic quark motion to the rotational motion as
an external perturbation, and consequently it is proportional to $\Omega$.
Here we confine ourselves to the mean field results
$( O (\Omega^0)$ contribution) to the above GPDF.
This leading term contributes to the isoscalar 
combination of $H_E (x, \xi, t)$, while it contributes to the isovector 
combination of $E_M(x, \xi,t)$ :
\begin{eqnarray}
 H_E^{(I = 0)} (x, \xi, t) &=& M_N N_c \int \frac{d z^0}{2 \pi}
 \sum_{n \leq 0} e^{i z^0 (x M_N - E_n)} \int d^3 
 \mbox{\boldmath $x$} \nonumber \\
 &\times& \Phi^{\dagger}_n (\mbox{\boldmath $x$}) (1 + \gamma^0 \gamma^3)
 e^{- \,i \,(z_0/2) \,\hat{p}_3} 
 e^{i \mbox{\boldmath $\Delta$} \cdot \mbox{\boldmath $x$}} 
 e^{- \,i \,(z_0/2) \,\hat{p}_3} 
 \Phi_n (\mbox{\boldmath $x$}) , \label{gpdhe} \\
 E_M^{(I = 1)} (x, \xi, t) 
 &=& \frac{2 i M_N^2 N_c}{3 (\mbox{\boldmath $\Delta$}^{\bot})^2} 
 \int \frac{d z^0}{2 \pi}
 \sum_{n \leq 0} e^{i \cdot z^0 (x M_N - E_n)} \int d^3 
 \mbox{\boldmath $x$} \nonumber \\
 &\times& \Phi^{\dagger}_n (\mbox{\boldmath $x$}) 
 (1 + \gamma^0 \gamma^3)
 (\mbox{\boldmath $\tau$} \times \mbox{\boldmath $\Delta$})^3 
 e^{- \,i \,(z_0/2) \,\hat{p}_3} 
 e^{i \mbox{\boldmath $\Delta$} \cdot \mbox{\boldmath $x$}} 
 e^{- \,i \,(z_0/2) \,\hat{p}_3} 
 \Phi_n (\mbox{\boldmath $x$}) . \label{gpdem}
\end{eqnarray}
As shown by several previous papers, the first moments of the above GPDF
reduce to the following forms \cite{OPSUG04}-\cite{SBR02} :
\begin{eqnarray}
 \int_{-1}^1 H_E^{(I = 0)} (x, \xi, t) d x &=& \int d^3 x \,
 e^{i \mbox{\boldmath $\Delta$} \cdot \mbox{\boldmath $x$}} N_c \sum_{n \leq 0}  \Phi_n^{\dagger} (\mbox{\boldmath $x$}) \Phi_n (\mbox{\boldmath $x$}) \\ 
 \int_{-1}^1 E_M^{(I = 1)} (x, \xi, t) d x 
 &=& \frac{2 i M_N N_c}{3 (\mbox{\boldmath $\Delta$}^{\bot})^2} 
 \int d^3 \mbox{\boldmath $x$} \nonumber \\ 
 &\times& \sum_{n \leq 0} 
 \Phi^{\dagger}_n (\mbox{\boldmath $x$}) 
 (1 + \gamma^0 \gamma^3) 
 (\mbox{\boldmath $\tau$} \times 
 \mbox{\boldmath $\Delta$})^3 
 e^{i \mbox{\boldmath $\Delta$} \cdot \mbox{\boldmath $x$}} 
 \Phi_n (\mbox{\boldmath $x$}) .
\end{eqnarray}
Here the symbol $\sum_{n \leq 0}$ denotes the summation over the
occupied (the valence plus negative-energy Dirac sea) single-quark
orbitals in the hedgehog mean field.
These expressions are slightly different from the ones given in several 
previous studies \cite{OPSUG04}-\cite{GPV01}.
In these studies, on the basis of the large $N_c$ 
argument, the l.h.s. of (\ref{hecomb}) is replaced by
$H^{(I = 0)} (x, \xi, t)$, 
while l.h.s. of (\ref{emcomb}) by $E^{(I = 1)} (x, \xi, t)$,
since the remaining terms are subleasing in $N_c$. 
Here, we retain these subleading terms because of 
the reason explained shortly.
The r.h.s. of these equations are the known theoretical expressions of
the Sachs form factor and the isovector magnetic form factor within
the CQSM, i.e.
\begin{eqnarray}
 \int_{-1}^1 H_E^{(I = 0)} (x, \xi, t) d x &=& G_E^u (t) +  G_E^d (t)
 = 3 G_E^{(I = 0)} (t) , \\
 \int_{-1}^1 E_M^{(I = 1)} (x, \xi, t) d x &=& G_M^u (t) -  G_M^d (t)
 =  G_M^{(I = 1)} (t) .
\end{eqnarray}
The reason why we did not drop the subleading terms $N_c$ 
in the l.h.s. of (\ref{hecomb}) and (\ref{emcomb}) is as follows.
If we did so, we would have obtained the 
relations
\begin{eqnarray}
 \int_{-1}^1 H^{(I = 0)} (x, \xi, t) d x &=& 3 \,G_E^{(I = 0)} (t) , \\
 \int_{-1}^1 E^{(I = 1)} (x, \xi, t) d x &=& G_M^{(I = 1)} (t) ,
\end{eqnarray}
which contradicts the first moment sum rule
expected on the general ground, i.e.
\begin{eqnarray}
 \int_{-1}^1 H^{(I = 0)} (x, \xi, t) d x &=& 3 \,F_1^{(I = 0)} (t) , \\
 \int_{-1}^1 E^{(I = 1)} (x, \xi, t) d x &=& F_2^{(I = 1)} (t) .
\end{eqnarray}
The first case would make little difference, because the difference
between $F_1^{(I = 0)} (t)$ and $G_E^{(I = 0)} (t)$ is small under
the circumstance in which the soliton center-of-mass motion is
nonrelativistic. This is not the case with the 
difference between $F_2^{(I = 1)} (t)$ and $G_M^{(I = 1)} (t)$,
as seen from the experimentally known relations :
\begin{eqnarray}
 F_2^{(I = 1)} (t = 0) &\simeq& \kappa^p - \kappa^n \simeq 3.7 ,\\
 G_M^{(I = 1)} (t = 0) &\simeq& 1 + (\kappa^p - \kappa^n)  \simeq 4.7 ,
\end{eqnarray}
although the anomalous magnetic moment term dominates over the Dirac
moment term both from the viewpoint of the $N_c$ counting as well as
numerically.
Next, we consider the forward limit of (\ref{gpdhe}) and (\ref{gpdem}).
The forward limit of (\ref{gpdhe}) gives
\begin{eqnarray}
 H_E^{(I = 0)} (x, 0, 0) &=& M_N N_c \int \frac{d z^0}{2 \pi}
 \sum_{n \leq 0} \,e^{i z^0 (x M_N - E_n)} \nonumber \\
 &\times& \int d^3 \mbox{\boldmath $x$} \Phi_n^{\dagger} 
 (\mbox{\boldmath $x$})
 (1 + \gamma^0 \gamma^3) e^{-i z_0 \hat{p}_3} 
 \Phi_n (\mbox{\boldmath $x$}) \nonumber \\
 &=& M_N N_c \sum_{n \leq 0} \langle n | (1 + \gamma^0 \gamma^3)
 \delta (x M_N - E_n - \hat{p}_3) | n \rangle .
\end{eqnarray}
The r.h.s. precisely coincides with the expression of the isoscalar
unpolarized quark distribution in the CQSM \cite{PPPBGW98},
i.e. we confirm that 	
\begin{equation}
 H_E^{(I = 0)} (x, 0, 0) = f_1^{(I = 0)} (x) .
\end{equation}
It is already known that the model expression for $f_1^{(I = 0)} (x)$
satisfies the quark number and the energy momentum sum
rules \cite{DPPPW96},\cite{DPPPW97} :
\begin{eqnarray}
 \int_{-1}^1 \,f_1^{(I = 0)} (x) d x &=& 3 , \\
 \int_{-1}^1 \,x \,f_1^{(I = 0)} (x) d x &=& 1 .
\end{eqnarray}
The forward limit of (\ref{gpdem}) reduces to
\begin{eqnarray}
 E_M^{(I = 1)} (x, 0, 0) 
 &=& \frac{1}{3} M_N^2 N_c \int \frac{d z^0}{2 \pi} \sum_{n \leq 0}
 e^{i z_0 (x M_N - E_n)} \nonumber \\
 &\times& \int d^3 \mbox{\boldmath $x$} \Phi_n^{\dagger} 
 (\mbox{\boldmath $x$}) 
 (\hat{\mbox{\boldmath $x$}} \times 
 \mbox{\boldmath $\tau$})_3 (1 + \gamma^0 \gamma^3) 
 e^{-i z_0 \hat{p}_3} \Phi_n (\mbox{\boldmath $x$}) \nonumber \\
 &=& \frac{1}{3} M_N^2 \cdot N_c \sum_{n \leq 0} \langle n 
 | (\hat{\mbox{\boldmath $x$}} \times 
 \mbox{\boldmath $\tau$})_3 (1 + \gamma^0 \gamma^3) 
 \delta (x M_N - E_n - \hat{p}_3) | n \rangle . \label{emforwardocc}
\end{eqnarray}
The first moment of this quantity gives 
\begin{equation}
 \int_{-1}^1 E_M^{(I = 1)} (x, 0, 0) d x = -\frac{M_N}{9} N_c
 \sum_{n \leq 0} \langle n | (\hat{\mbox{\boldmath $x$}} \times 
 \mbox{\boldmath $\alpha$}) \cdot \mbox{\boldmath $\tau$}
 | n \rangle . \label{mom1st}
\end{equation}
The r.h.s. precisely gives the theoretical expression for the isovector 
magnetic moment in the CQSM.  (It is not the anomalous magnetic
moment part.) This is basically a known fact \cite{PPPBGW98}, but what we
emphasize here is that $E_M^{(I = 1)} (x, 0, 0)$ is interpreted
to give the distribution of (isovector) nucleon magnetic moment in the 
Feynman $x$ space not in the ordinary coordinate space.

Now, we turn to our main concern in this paper, i.e.
the second moment of this quantity, which we expect is related to
the isovector part of the quark angular momentum fraction of
the nucleon.
Performing a weighted $x$ integral, we find that
\begin{equation}
 \int_{-1}^1 x E_M^{(I = 1)} (x, 0, 0) d x
 \ = \ \frac{1}{3} N_c \sum_{n \leq 0} \langle n | 
 (\hat{\mbox{\boldmath $x$}}
 \times \mbox{\boldmath $\tau$})_3 (1 + \alpha_3) 
 (E_n + \hat{p}_3) | n \rangle .
\end{equation}
One notices that the second moment of $E_M^{(I = 1)} (x, 0, 0)$ can be 
decomposed into four parts as
\begin{eqnarray}
 \int _{-1}^1 x E_M^{(I = 1)} (x, 0, 0) d x
 &=& \frac{1}{3} N_c \sum_{n \leq 0} 
 \langle n | (\hat{\mbox{\boldmath $x$}} \times 
 \mbox{\boldmath $\tau$})_3 E_n | 
 n \rangle \ + \ 
 \frac{1}{3} N_c \sum_{n \leq 0} 
 \langle n | (\hat{\mbox{\boldmath $x$}} \times 
 \mbox{\boldmath $\tau$})_3 \alpha_3 
 E_n | n \rangle \nonumber \\
 &+& \frac{1}{3} N_c \sum_{n \leq 0} 
 \langle n | (\hat{\mbox{\boldmath $x$}} \times 
 \mbox{\boldmath $\tau$})_3 
 \hat{p}_3 | n \rangle \ + \ 
 \frac{1}{3} N_c \sum_{n \leq 0} 
 \langle n | (\hat{\mbox{\boldmath $x$}} \times 
 \mbox{\boldmath $\tau$})_3 
 \alpha_3 \hat{p}_3 | n \rangle \nonumber \\
 &\equiv& M_1 + M_2 + M_3 + M_4 .
\end{eqnarray}
Using the Dirac equation (here $\hat{\mbox{\boldmath $r$}} \equiv
\hat{\mbox{\boldmath $x$}} / |\hat{\mbox{\boldmath $x$}}|$)
\begin{eqnarray}
 H | n \rangle &=& E_n | n \rangle ,
\end{eqnarray}
with
\begin{eqnarray}
 H &=& \mbox{\boldmath $\alpha$} \cdot \hat{\mbox{\boldmath $p$}}
  + M \beta 
 (\cos F(r) + i \gamma_5 \mbox{\boldmath $\tau$} \cdot 
 \hat{\mbox{\boldmath $r$}} \sin F(r) ) ,
\end{eqnarray}
the first term can be rewritten as 
\begin{eqnarray}
 M_1 &=& \frac{N_c}{3} \sum_{n \leq 0} \frac{1}{2}
 \langle n | \{ (\hat{\mbox{\boldmath $x$}} \times 
 \mbox{\boldmath $\tau$})_3, H \} | n \rangle \nonumber \\
 &=& \frac{N_c}{3} \sum_{n \leq 0} \frac{1}{2}
 \langle n | -i (\mbox{\boldmath $\alpha$} \times 
 \mbox{\boldmath $\tau$})_3 - 2(\mbox{\boldmath $\tau$} 
 \times \hat{\mbox{\boldmath $x$}})_3 \mbox{\boldmath $\alpha$} 
 \cdot \hat{\mbox{\boldmath $p$}} \,| n \rangle .
\end{eqnarray}
It is easy to see that this term vanishes identically due to the hedgehog 
symmetry. Alternatively, we can use the parity symmetry to show
$M_1 = 0$. The parity also enforces the fourth term $M_4$ to vanish.
In fact, under the parity operation ${\cal P}$, we have
\begin{eqnarray}
 {\cal P} | n \rangle &=& (-1)^{P_n} | n \rangle \\
 {\cal P} (\hat{\mbox{\boldmath $x$}} \times 
 \hat{\mbox{\boldmath $\tau$}})_3 {\cal P}^{-1}
 &=& -(\mbox{\boldmath $x$} \times \mbox{\boldmath $\tau$})_3, \\
 {\cal P} \alpha_3 {\cal P}^{-1} &=& -\alpha_3, \\
 {\cal P} \hat{p}_3 {\cal P}^{-1} &=& -\hat{p}_3
\end{eqnarray}
with $(-1)^{P_n}$ being the parity of the eigenstate $| n \rangle$,
so that we conclude that
\begin{eqnarray}
 M_4 &=& \frac{N_c}{3} \sum_{n \leq 0}
 \langle n | {\cal P}^{-1} {\cal P} 
 {(\hat{\mbox{\boldmath $x$}} \times \mbox{\boldmath $\tau$})}_3
 {\cal P}^{-1} {\cal P} \alpha_3 {\cal P}^{-1} {\cal P} 
 \hat{p}_3 {\cal P}^{-1} {\cal P} | n \rangle \nonumber \\
 &=& \{ (-1)^{P_n} \}^2  (-1)^3 \,M_4 \ = \ - \,M_4 \ = \ 0 .
\end{eqnarray}
The third term $M_3$ does not vanish but it can be simplified in the
following way because of the hedgehog symmetry 
(generalized spherical symmetry) as : 
\begin{eqnarray}
 M_3 &=& \frac{N_c}{3} \sum_{n \leq 0}
 \langle n | (\hat{\mbox{\boldmath $x$}} \times 
 \mbox{\boldmath $\tau$})_3 \,\hat{p}_3 
 | n \rangle
 \ = \ \frac{N_c}{3} \sum_{n \leq 0} \frac{1}{3}
 \langle n | (\hat{\mbox{\boldmath $x$}} \times 
 \mbox{\boldmath $\tau$}) \cdot  \hat{\mbox{\boldmath $p$}} 
 | n \rangle \nonumber \\
 &=& \frac{N_c}{3} \sum_{n \leq 0} \left(- \frac{1}{3}\right)
 \langle n |  \mbox{\boldmath $\tau$} \cdot 
 (\hat{\mbox{\boldmath $x$}} \times 
 \hat{\mbox{\boldmath $p$}}) | n \rangle
 \ = \ -\frac{N_c}{3} \sum_{n \leq 0} 
 \langle n | \tau_3 L_3 | n \rangle .
\end{eqnarray}
Finally, the second term $M_2$ can be rewritten in the following manner
by using the Dirac equation, commutation relations of $\gamma$-matrices
and isospin matrices, and also the hedgehog symmetry :
\begin{eqnarray}
  M_2 &=& \frac{N_c}{3} \sum_{n \leq 0} \frac{1}{2}
 \langle n |  \{ (\hat{\mbox{\boldmath $x$}} 
 \times \mbox{\boldmath $\tau$})_3 \alpha_3, 
 H \} | n \rangle \nonumber \\ 
 &=& \frac{N_c}{3} \sum_{n \leq 0} \frac{1}{2}
 \langle n |  \{ (\hat{\mbox{\boldmath $x$}} 
 \times \mbox{\boldmath $\tau$})_3 \alpha_3, 
 \mbox{\boldmath $\alpha$} \cdot\hat{\mbox{\boldmath $p$}} \} | 
 n \rangle \nonumber \\
 &+& \frac{N_c}{3} \sum_{n \leq 0} \frac{1}{2}
 \langle n |  \{ (\hat{\mbox{\boldmath $x$}} 
 \times \mbox{\boldmath $\tau$})_3 \alpha_3, 
 M \beta (\cos F(r) + \gamma_5 \mbox{\boldmath $\tau$} 
 \cdot \hat{\mbox{\boldmath $r$}}
 \sin F(r) ) \} | n \rangle \nonumber \\ 
 &=& -\frac{N_c}{3} \sum_{n \leq 0} 
 \langle n | \tau_3 L_3 + \tau_3 \Sigma_3 | n \rangle \nonumber \\
 &\,& - \,M \cdot \frac{N_c}{9} \sum_{n \leq 0} 
 \langle n | r \sin F(r) \gamma^0 [ \mbox{\boldmath $\Sigma$} 
 \cdot \hat{\mbox{\boldmath $r$}}
 \mbox{\boldmath $\tau$} \cdot \hat{\mbox{\boldmath $r$}} 
 - \mbox{\boldmath $\Sigma$} \cdot 
 \mbox{\boldmath $\tau$} ] | n \rangle . 
\end{eqnarray}
Collecting all the four terms, we finally obtain the second moment sum 
rule of the form 
\begin{eqnarray}
 \int_{-1}^1 x E_M^{(I = 1)} (x, 0, 0) d x
 &=& 2 \,\left(-\frac{N_c}{3} \right) \sum_{n \leq 0}
 \langle n | \tau_3 
 \left(L_3 + \frac{1}{2} \Sigma_3 \right) | n \rangle \nonumber \\
 &\,& - \,M \cdot \frac{N_c}{9} \sum_{n \leq 0}
 \langle n | r \sin F(r) \gamma^0 [ \mbox{\boldmath $\Sigma$} 
 \cdot \hat{\mbox{\boldmath $r$}}
 \mbox{\boldmath $\tau$} \cdot \hat{\mbox{\boldmath $r$}}
 - \mbox{\boldmath $\Sigma$} \cdot 
 \mbox{\boldmath $\tau$} ] | n \rangle .
\end{eqnarray}
The first term of the r.h.s. just coincides with the proton matrix 
element of the free field quark angular momentum operator, or more 
precisely its isovector part, given by
\begin{equation}
 J_f^{(I = 1)} = \langle p \uparrow | \hat{J}_f^{(I = 1)} 
 | p \uparrow \rangle ,
\end{equation}
with
\begin{eqnarray}
 \hat{J}_f^{(I = 1)} &\equiv& \int 
 {\psi}^\dagger (\hat{\mbox{\boldmath $x$}}) \,\tau_3
 \,\left[ (\mbox{\boldmath $x$} \times \hat{\mbox{\boldmath $p$}})_3 
 + \frac{1}{2} \Sigma_3 \right] \,
 \psi (\mbox{\boldmath $x$}) \,d^3 x
 \ = \ \hat{L}_f^{(I = 1)} + \frac{1}{2} \,\hat{\Sigma}^{(I = 1)} .
\end{eqnarray}
In view of Ji's angular momentum sum rule,
we would have naively expected 
to get  
\begin{equation}
 \frac{1}{2} \,\int_{-1}^1 x E_M^{(I = 1)} (x, 0, 0) d x = J_f^{(I = 1)} .
\end{equation}
Somewhat unexpectedly, however, we find that
\begin{eqnarray}
 \frac{1}{2} \int_{-1}^1 x E_M^{(I = 1)} (x, 0, 0) d x
 \ = \ J_f^{(I = 1)} + \delta J^{(I = 1)} , \label{emiv2nd}
\end{eqnarray}
with
\begin{equation}
 \delta J^{(I = 1)} = -M \,\frac{N_c}{18} \sum_{n \leq 0}
 \langle n | r \sin F(r) \gamma^0 
 [ \mbox{\boldmath $\Sigma$} \cdot 
 \hat{\mbox{\boldmath $r$}} \mbox{\boldmath $\tau$} 
 \cdot \hat{\mbox{\boldmath $r$}} - \mbox{\boldmath $\Sigma$} 
 \cdot \mbox{\boldmath $\tau$} ] | n \rangle .
\end{equation}
This second moment sum rule should be contrasted with the corresponding 
sum rule in the isoscalar channel, which was recently proved
in \cite{OPSUG04} :
\begin{equation}
 \frac{1}{2} \int_{-1}^1 x E_M^{(I = 0)} (x, 0, 0) \,d x \ = \ 
 L_f^{(I=0)} + \frac{1}{2} \Delta \Sigma^{(I=0)} 
 \ = \ J_f^{(I = 0)} \ = \ 1/2 , \label{emis2nd}
\end{equation}
which appears to be compatible with Ji's general sum rule, considering
that the nucleon spin sum rule should be saturated by the 
quark field alone in the CQSM \cite{WY91}. How should we interpret the
above unexpected result for the isovector case ?
One may argue that the second moment need not necessarily reduce to the 
nucleon matrix element of the free field angular momentum operator 
since the CQSM is anyhow an interacting theory of quark fields.
In fact, the above derivation of the sum rule (\ref{emiv2nd})
indicates that the cause of the $\delta J^{(I = 1)}$ term
can be traced back to the dynamical 
generation of the effective quark mass and the formation of the 
symmetry breaking mean field containing the scalar product of
$\mbox{\boldmath $\tau$}$ and $\hat{\mbox{\boldmath $r$}}$.
At least, one can say that, since
the $\delta J^{(I = 1)}$ term is proportional to the dynamically
generated quark mass $M$, it vanishes in the 
perturbative vacuum, although it is meaningless to consider the chiral 
soliton if the QCD vacuum is perturbative.
We recall that a similar breakdown of the second moment sum rule, expected 
on the general ground of QCD, occurs also in the case of chiral-odd 
twist-3 distribution functions $e(x)$ of the nucleon. From the general 
QCD analysis, one expect that the second moment of the isoscalar part of 
$e(x)$ satisfies the following sum rule in the chiral
limit \cite{ES03},
\begin{equation}
 \int_{-1}^1 x \,e^{(I = 0)} (x) d x = 0 .
\end{equation}
In the CQSM, however, we obtain \cite{S03},\cite{OW04}
\begin{equation}
 \int_{-1}^1 x \,e^{(I = 0)} (x) d x 
 = \frac{M}{M_N} N_c \sum_{n \leq 0}
 \langle n | \frac{1}{2} \left( U + U^\dagger \right) | n \rangle ,
\end{equation}
with $U = e^{i \mbox{\boldmath $\tau$} \cdot \hat{\mbox{\boldmath $r$}}
F(r)}$. It is indicative that the violation of the QCD sum rule is again 
proportional to the dynamical quark mass $M$, which would vanish 
in the perturbative vacuum \cite{OW04}.

From the practical viewpoint, only the l.h.s. of
(\ref{emiv2nd}) is observable, and neither of $J_f^{(I = 1)}$ nor
$\delta J^{(I = 1)}$ is observable. We therefore take the following
viewpoint, although it is not absolutely mandatory.
The second moment of $E_M^{(I = 1)} (x,0, 0)$ gives the isovector quark 
angular momentum fraction of the interacting theory (it is the CQSM 
in the present context) :
\begin{equation}
 \frac{1}{2} \int_{-1}^1 x E_M^{(I = 1)} (x, 0, 0) d x 
 \equiv J^{(I = 1)} \equiv J^u - J^d .
\end{equation}
Although somewhat arbitrary, this sum rule combined with
(\ref{emis2nd}) allows us to carry out the flavor decomposition
of the quark angular momentum fraction in the nucleon.
In any case, the most important quantity here is 
$E_M^{(I = 1)} (x, 0, 0)$, since it is the quantity which can
be directly measured through the DVCS experiments.
In the next section, we perform a numerical calculation of 
$E_M^{(I = 1)} (x, 0, 0)$ without recourse to the derivative expansion 
type approximation. We also evaluate $J_f^{(I = 1)}$ and 
$\delta J^{(I = 1)}$ in the r.h.s. of (\ref{emiv2nd})
directly without any notion of distribution functions.
This allows us to check the validity of the sum rule (\ref{emiv2nd})
within the model, thereby providing us with a nontrivial
check for our numerical result for $E_M^{(I = 1)} (x, 0, 0)$.

\section{Numerical results and discussion}

 The most important parameter of the CQSM is the dynamical quark
mass $M$, which plays the role of the quark-pion coupling constant,
thereby controlling basic soliton
properties \cite{DPP88},\cite{WY91}. Here we use the value
$M = 375 \,\mbox{MeV}$, which is favored from the analysis of nucleon
low energy observables. 
The model is an effective theory which is defined with an physical
cutoff. We use here the double-subtraction Pauli-Villars regularization
scheme proposed in \cite{KWT99}. (Naturally, this is not the only
way of introducing regularization. More sophisticated regularization
scheme is proposed in \cite{WAG99}.)
In this scheme, the most nucleon observables are regularized 
through the subtraction.
\begin{equation}
 \langle O \rangle^{reg} \equiv \langle O \rangle^M - \sum_{i = 1}^2 c_i 
 \langle O \rangle^{\Lambda_i} .
\end{equation}
Here $\langle O \rangle^M$ denotes the nucleon matrix element of an 
operator $O$ evaluated with the original action of the CQSM having the
mass parameter $M$, while $\langle O \rangle ^{\Lambda_i}$ stands for the 
corresponding matrix element obtained from $\langle O \rangle^M$ by 
replacing the parameter $M$ with the Pauli-Villars cutoff mass
$\Lambda_i$. To remove all the ultraviolet divergences of the theory,
the parameters $c_i$ and $\Lambda_i$ must satisfy the conditions :
\begin{eqnarray}
 M^2 - \sum_{i = 1}^2 c_i \Lambda_i^2 &=& 0 , \label{pvcond1} \\
 M^4 - \sum_{i = 1}^2 c_i \Lambda_i^4 &=& 0 . \label{pvcond2}
\end{eqnarray}
We further impose two additional physical conditions, which amounts to 
requiring that the model reproduces the empirically known value of the
vacuum quark condensate as well as the correct normalization of the
pion kinetic term in the corresponding bosonized action. This gives
\begin{eqnarray}
 \langle \bar{\psi} \psi \rangle_{vac} &=& \frac{N_c M^3}{2 \pi^2}
 \sum_{i = 1}^2 c_i \left(\frac{\Lambda_i}{M}\right)^4 
 \ln \left(\frac{\Lambda_i}{M}\right)^2 , \label{pvcond3} \\
 f_{\pi}^2 &=& \frac{N_c M^2}{4 \pi^2} \sum_{i = 1}^2 c_i
 \left(\frac{\Lambda_i}{M}\right)^2 \ln 
 \left(\frac{\Lambda_i}{M}\right)^2 . \label{pvcond4}
\end{eqnarray}
These four conditions (\ref{pvcond1}), (\ref{pvcond2}), 
(\ref{pvcond3}) and (\ref{pvcond4}) are enough to fix the four 
parameters $c_1, c_2, \Lambda_1$ and $\Lambda_2$. 
Fixing $M$ and $f_{\pi}$ to be 375MeV and 93 MeV, respectively,
we find
\begin{eqnarray}
 c_1 &=& 0.39937, \ \ \ c_2 = -0.00661, \\
 \Lambda_1 &=& 627.653 \,\mbox{MeV}, \ \ \ 
 \Lambda_2 = 1589.45 \,\mbox{MeV} .
\end{eqnarray}
As usual, the numerical calculations in the CQSM are carried out 
by using Kahana-Ripka's discretized momentum
basis \cite{KR84},\cite{KRS84}. Our main task here is 
to evaluate $E_M^{(I = 1)} (x, 0, 0)$ given by (\ref{emforwardocc})
making use of the Kahana-Ripka basis, which turns out not so easy.
The cause of difficulty lies in the 
following fact. First, the appearance of the momentum operator
$\hat{p}_3$ in the delta function enforces us to work in the momentum
representation. On the other hand, the matrix element
in (\ref{emforwardocc}) contains the coordinate 
operator $\hat{\mbox{\boldmath $x$}}$. This coordinate operator becomes a 
differential operator in the momentum representation, which is however
incompatible with the whole calculation scheme of the CQSM making use 
of the discretized momentum basis. To circumvent this difficulty, we
insert a complete set of states $| m_0 \rangle$ as follows :
\begin{eqnarray}
 E_M^{(I = 1)} (x, 0, 0) &=& \frac{1}{3} M_N^2 \cdot N_c 
 \sum_{n \leq 0} \sum_{m_0 = all} \nonumber \\
 &\times& \langle n | (\hat{\mbox{\boldmath $x$}} \times 
 \mbox{\boldmath $\tau$})_3
 (1 + \gamma^0 \gamma^3) | m_0 \rangle \langle m_0 | \delta 
 (x M_N - E_n - \hat{p}_3) | n \rangle  \label{compset}.
\end{eqnarray}
This complete set can in principle be chosen at will.
It can be the eigenstates of
the full Dirac Hamiltonian $H$ as the states $| n \rangle$ are so, or it
can be the eigenstates of the free Dirac Hamiltonian $H_0$ given by
\begin{equation}
 H_0 = \mbox{\boldmath $\alpha$} \cdot \hat{\mbox{\boldmath $p$}}
 + \beta M .
\end{equation}
We choose here the latter. The grand spin $K^{\prime}$ of the states 
$| m_0 \rangle$ need not be the same as the grand spin $K$ of the
states $| n \rangle$, 
but the finite rank nature of the operator 
$(\hat{\mbox{\boldmath $x$}} \times \mbox{\boldmath $\tau$})_3 
(1 + \gamma^0 \gamma^3)$ 
restricts the value of $K^{\prime}$ to be $K$, $K \pm 1$, or $K \pm 2$.
The advantage of the expression (\ref{compset}) is that the first
and the second matrix elements can  respectively be evaluated in the
coordinate representation and the momentum representation.

The theoretical expression for $E_M^{(I=1)}(x,0,0)$
in (\ref{emforwardocc}) is given as the summation over the occupied
single-quark orbitals. An alternative but equivalent expression
is obtained for it, which is given as the summation over the
nonoccupied quark levels as
\begin{equation}
 E_M^{(I = 1)} (x, 0, 0) 
 \ = \ - \frac{1}{3} M_N^2 \cdot N_c \sum_{n > 0} \langle n 
 | (\hat{\mbox{\boldmath $x$}} \times 
 \mbox{\boldmath $\tau$})_3 (1 + \gamma^0 \gamma^3) 
 \delta (x M_N - E_n - \hat{p}_3) | n \rangle . \label{emforwardnocc}
\end{equation}
The equivalence of these two representations is based on a quite
general principle of field theory, i.e. the anticommuting property
of two quark field operator with space-like separation.
It is also known that the Pauli-Villars regularization preserves
this equivalence. For numerical calculation of $E_M^{(I=1)}(x,0,0)$,
it is convenient to use the occupied form (\ref{emforwardocc}) for
$x > 0$, and the nonoccupied form (\ref{emforwardnocc}) for $x < 0$.
(We recall that the distribution with $x < 0$ is related to the
antiquark distribution with $x > 0$.)

\begin{figure}[htb] \centering
\begin{center}
 \includegraphics[width=11.0cm,height=9.5cm]{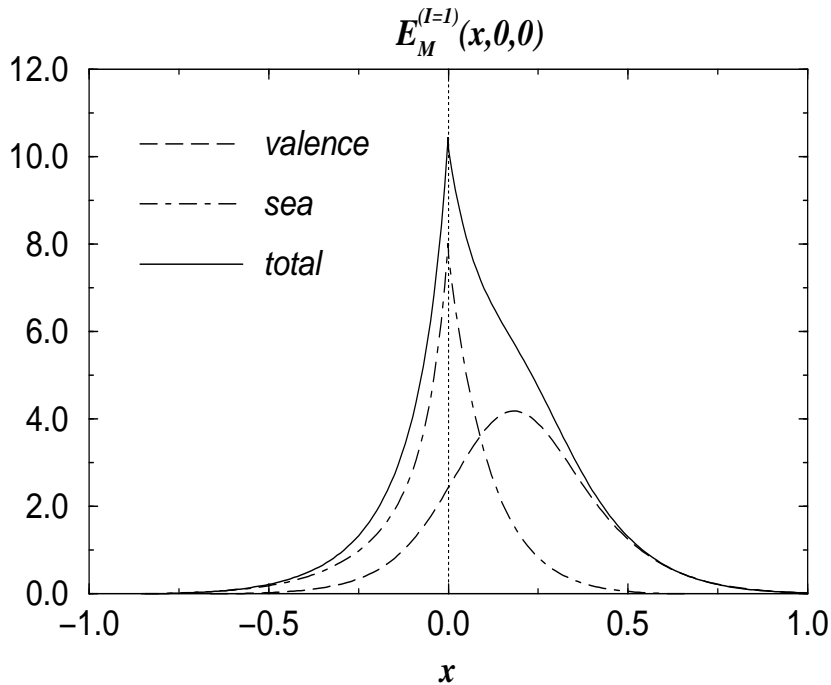}
\end{center}
\vspace*{-0.5cm}
\renewcommand{\baselinestretch}{1.20}
\caption{The CQSM prediction for $E_M^{(I=1)}(x,0,0)$.
The long-dashed and dash-dotted curves here stand for the contribution
of the valence quarks and of the Dirac sea quarks, while their
sum is represented by the solid curve.}
\label{Fig:gpdf}
\end{figure}

Now, we show our numerical result for $E_M^{(I = 1)} (x, 0, 0)$. The 
long-dashed and the dash-dotted curves in Fig.\ref{Fig:gpdf} stand for the
contributions of the three valence quarks and of the Dirac sea quarks,
respectively.
The sum of these two contributions is represented by the solid curve.
A remarkable feature here is that the contribution of the Dirac sea
quarks has a sharp peak around $x = 0$. This confirms the
qualitative result first given in \cite{GPV01}.
Here we recall the fact that the $x$-integral or the first moment of
$E_M^{(I = 0)} (x, 0, 0)$ gives the isovector magnetic moment
$\mu^{(I = 1)} = \mu_p -\mu_n$ of the nucleon.
This denotes that $E_M^{(I = 1)} (x, 0, 0)$ gives the distribution of the 
nucleon isovector magnetic moment in the Feynman $x$ space not in the
ordinary coordinate space.
The sharp peak around $x = 0$ therefore means that 
the quark and antiquark with small $x$ carry a sizable amount of isovector
magnetic moment of the nucleon.
We interpret this fact as an indication of the importance of the quark
motion in the transverse direction by the following reason.
Firstly, the isovector nucleon magnetic moment mainly comes from the
quark and antiquark orbital motion in the nucleon, since it is known that
the anomalous part dominate over the Dirac moment part.
Secondly, the quark and antiquark with $x = 0$ has zero velocity in the 
longitudinal direction. Accordingly, the large magnetic moment density
concentrated in the small $x$ region must come from the motion of
quark and antiquarks in the plane perpendicular to the proton spin
direction.

The prominent peak of $E_M^{(I=1)}(x,0,0)$ around $x=0$ can also be
interpreted as the effect of pionic $q \bar{q}$ excitation with
large spatial extension. In fact, it has long been known that the
pion cloud around the ``bare'' nucleon gives a significant
contribution to the isovector nucleon magnetic moment.
This then indicates that the dominant contribution to
$E_M^{(I=1)}(x,0,0)$ in the small $x$ domain originates from the
motion of correlated quarks and antiquarks, the spatial distribution
of which have a long range tail in the transverse direction.
The validity of the proposed interpretation may be tested more
definitely, if one could evaluate the so-called impact-parameter
dependent distribution function defined by \cite{B00}\nocite{B03}
\nocite{RJB00}\nocite{RP02}--\cite{BJY04}.
\begin{equation}
 \varepsilon_M^{(I=1)}(x,\mbox{\boldmath $b$}_\perp) = 
 \int \frac{d^2 \mbox{\boldmath $\Delta$}_\perp}{(2 \pi)^2} \,
 e^{- i \mbox{\boldmath $\Delta$}_\perp \cdot 
 \mbox{\boldmath $b$}_\perp} \,
 E_M^{(I=1)}(x,0,- \mbox{\boldmath $\Delta$}_\perp^2) .
\end{equation}

The rest of this section will be devoted to the numerical check of
the first and second moment sum rule of $E_M^{(I = 1)} (x, 0, 0)$ within
the framework of the CQSM. This is important not only to confirm the
precision of our numerical result for $E_M^{(I = 1)} (x, 0, 0)$ but
also to verify the internal consistency of the whole theoretical framework.
We first discuss the first moment sum rule given by (\ref{mom1st}).
The point is that both sides of this equation can be evaluated totally
independently. The l.h.s. can be is calculated numerically integrating
the already given $E_M^{(I = 0)} (x, 0, 0)$ over $x$.
On the other hand, the evaluation of the r.h.s., i.e. the isovector 
magnetic moment of the nucleon, has no trouble, since it is just
a nucleon expectation value of a local operator. Numerically, we have 
got 
\begin{equation}
 \int_{-1}^1 E_M^{(I = 1)} (x, 0, 0) \,d x 
 \ \simeq \ 2.07 + 1.79 \ \simeq \ 3.86 ,
\end{equation}
while
\begin{equation}
 - \frac{M_N}{9} \,N_c \,\sum_{n \leq 0} \,
 \langle n \vert \,(\hat{\mbox{\boldmath $x$}} \times 
 \mbox{\boldmath $\alpha$}) \cdot 
 \mbox{\boldmath $\tau$} \vert n \rangle
 \ = \ \mu_V^{(I = 1)} \ \simeq \ 2.05 + 1.87 \ \simeq \ 3.92 ,
\end{equation}
which coincides with the precision of about $1\%$.
One may notice that the theoretical prediction for $\mu_V^{(I = 1)}$
is smaller than the experimental value
$\mu_V^{(I = 1)} (\exp) \simeq 4.7$.
We however know that, within the framework of the CQSM, there is a
rotational correction to $\mu_V^{(I = 1)}$ proportional to the
collective angular velocity $\Omega$, which is known to fill
this gap \cite{WW94},\cite{CBGPPWW94}. (We should however recall
some controversy related to this first order rotational correction
to some isovector nucleon observables. \cite{AW93}\nocite{SW95A}
\nocite{SW95B}--\cite{W96}.)
This first order rotational correction is naturally expected to
contribute also to $E_M^{(I = 1)}(x, 0, 0)$,
thereby to both side of the first moment sum rule. The calculation of 
such a higher order contribution to $E_M^{(I = 1)}(x, 0, 0)$
is beyond the scope of the present paper.
Next we turn to the second moment sum rule of $E_M^{(I = 1)} (x, 0, 0)$
given by (\ref{emiv2nd}).
We first evaluate the first term of the r.h.s. of (\ref{emiv2nd}), i.e.
the nucleon matrix element of the free field angular momentum operator
in the isovector combination. This term consists of the two terms as 
\begin{equation}
 J_f^{(I = 1)} = L_f^{(I = 1)} + \frac{1}{2} \,\Delta \Sigma^{(I = 1)} .
\end{equation}
Here, $L_f^{(I = 1)}$ is the nucleon matrix element of the free field
isovector angular momentum operator, while 
$\Delta \Sigma^{(I = 1)}$ is the isovector part of the longitudinal quark 
polarization.

\vspace{3mm}
\newcommand{\lw}[1]{\smash{\lower2.ex\hbox{#1}}}
\begin{table}[h]
\caption{The separate contributions of the valence and the Dirac sea
quarks to the quantities $L_f^{(I=1)}$, $\Delta \Sigma^{(I=1)}$,
and $J_f^{(I=1)} \equiv \ \ L_f^{(I=1)} + \frac{1}{2} \,
 \Sigma^{(I=1)}$ defined in the text.}
\label{Table:spin}
\vspace{2mm}
\begin{center}
\renewcommand{\arraystretch}{1.0}
\begin{tabular}{cccc}
\hline\hline
 & \ \ \ \ $L_f^{(I=1)}$ \ \ \ & \ \ $\Sigma^{(I=1)}$ \ \ &
 \ \ $J_f^{(I=1)}$ \\
 \hline
 \ \ valence \ \ & 0.147 & 0.705 & 0.5000 \\
 \hline
 \ \ sea \ \ & -0.265 & 0.357 & -0.087 \\
 \hline
 \ \ total \ \ & -0.115 & 1.057 & 0.413 \\
\hline \hline
\end{tabular}
\end{center}
\end{table}

We show in Table \ref{Table:spin} the separate contribution of the
valence quarks and the Dirac sea quarks to
$L_f^{(I = 1)}, \Delta \Sigma^{(I = 1)}$ and $J_f^{(I = 1)}$.
One sees that the valence quark 
contribution to $J_f^{(I = 1)}$ is precisely 1/2. while the Dirac sea 
contribution to it is slightly negative.

\vspace{3mm}
\begin{table}[h]
\caption{The separate contributions of the valence and the Dirac sea
quarks to the quantities $J_f^{(I=1)}$, $\delta J^{(I=1)}$,
and their sum. Also shown are the corresponding numbers for
$\frac{1}{2} \,\int \,x \,E_M^{(I=1)} (x,0,0) \,dx$.}
\label{Table:moment}
\vspace{2mm}
\begin{center}
\renewcommand{\arraystretch}{1.0}
\begin{tabular}{ccccc}
\hline\hline
 & \ \ \ \ $J_f^{(I=1)}$ \ \ \ & \ \ $\delta J^{(I=1)}$ \ \ & 
 \ \ $J_f^{(I=1)} + \delta J^{(I=1)}$ \ \ & \ \ 
 $\frac{1}{2} \,\int \,x \,E_M^{(I=1)} (x,0,0) \,dx$ \ \ \\
 \hline
 \ \ valence \ \ & 0.500 & -0.289 & 0.211 & 0.210 \\
 \hline
 \ \ sea \ \ & -0.087 & 0.077 & -0.010 & -0.008 \\
 \hline
 \ \ total \ \ & 0.413 & -0.212 & 0.201 & 0.202 \\
\hline\hline
\end{tabular}
\end{center}
\end{table}

Next, shown in Table \ref{Table:moment} are the 
contributions of the valence quarks and the Dirac sea quarks to the
quantity $\delta J^{(I = 1)}$ in the second moment sum rule as well
as the sum of $J_f^{(I = 1)}$ and $\delta J^{(I = 1)}$.
(The numerical values of $J_f^{(I = 1)}$ already given in Table 1
are also shown for convenience.)
Also shown in this table is a half of the 
second moment of $E_M^{(I = 1)} (x, 0, 0)$ obtained numerically from
the weighted $x$ integral of it. One confirms that the second moment
sum rule (\ref{emiv2nd}) is satisfied with good precision,
which in turn assures the numerical 
accuracy of our numerical calculation of $E_M^{(I = 1)} (x, 0, 0)$.
Knowing that the second moment of $E_M^{(I = 1)} (x, 0, 0)$ does not
coincide with the nucleon matrix element of the free field quark
angular momentum operator in the isovector case,
we would rather take a viewpoint that its 
second moment gives the quark angular momentum in the interacting theory.
This amounts to regarding the sum of $J_f^{(I = 1)}$ and 
$\delta J^{(I = 1)}$ as the isovector part of the quark angular momentum 
fraction $J^{(I = 1)}$ of the interacting theory, which gives
\begin{equation}
 J^{(I = 1)} = J^u - J^d \simeq 0.202 .
\end{equation}
The corresponding quark orbital angular momentum fraction can be
estimated by subtracting the corresponding quark spin part 
$\frac{1}{2} \Delta \Sigma^{(I = 1)} \simeq 0.529$, which gives
\begin{equation}
L^{(I = 1)} = L^u - L^d \simeq -0.327 .
\end{equation}
Worthy of mention here is the $x$-distribution of the isovector quark
angular momentum. Pushing forward with the above interpretation, let us
identify $\frac{1}{2}x E_M^{(I = 0)} (x, 0, 0)$ with the isovector
quark angular momentum distribution
$J^{(I = 1)} (x) \equiv J^u (x) - J^d (x)$.

\begin{figure}[htb] \centering
\begin{center}
 \includegraphics[width=11.0cm,height=9.5cm]{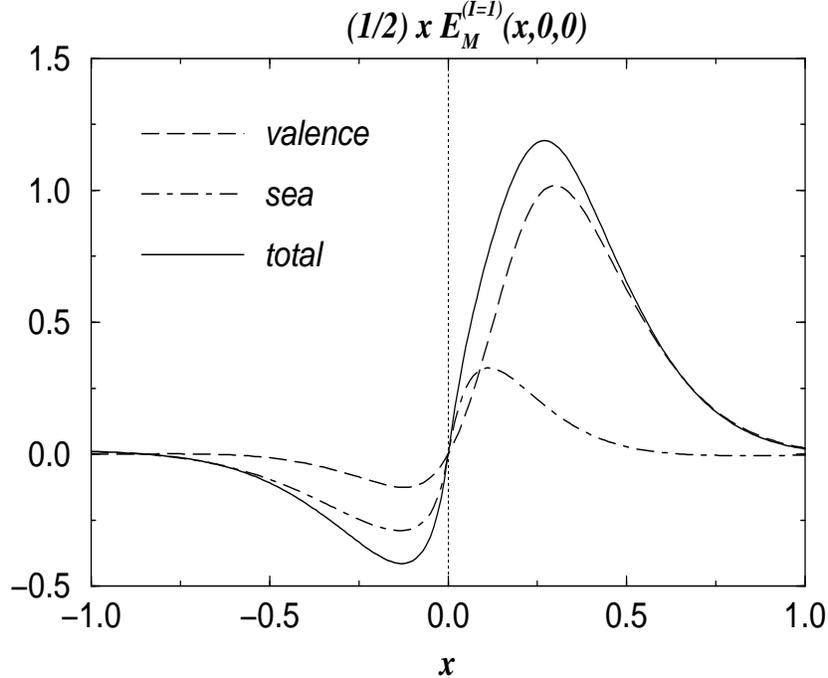}
\end{center}
\vspace*{-0.5cm}
\renewcommand{\baselinestretch}{1.20}
\caption{The CQSM prediction for $\frac{1}{2} \,x \,
E_M^{(I=1)}(x,0,0)$. The meaning of the curves is the same as
in Fig.\ref{Fig:gpdf}.}
\label{Fig:xgpdf}
\end{figure}%

Shown in Fig.~\ref{Fig:xgpdf} is the CQSM prediction for this quantity 
$\frac{1}{2}x E_M^{(I = 0)} (x, 0, 0)$, or $J^u (x) - J^d (x)$ in the
above interpretation.
Here, the distribution in the negative  region should be interpreted as 
that of antiquarks, i.e. $J^u (-x) - J^d (-x) = 
J^{\bar{u}} (x) - J^{\bar{d}} (x)$ with $x > 0$. 
Then, we observe from this figure that 
\begin{equation}
 J^u (x) - J^d (x) > 0 \ \ \ (\mbox{for} \ x > 0) ,
\end{equation}
while
\begin{equation}
 J^{\bar{u}} (x) - J^{\bar{d}} (x) < 0 \ \ \ (\mbox{for} \ x > 0) .
\end{equation}
The first inequality is nothing surprising, since the proton contains 
two $u$-quarks and one $d$-quark as valence particles.
More interesting here is the second inequality, which indicates that
the $\bar{d}$-quark carries more 
angular momentum than the $\bar{u}$-quark in the proton.
This reminds us of the violation of the Gottfried sum rule, which
has been accepted by now as a clear evidence of the dominance of the
$\bar{d}$-quark over the $\bar{u}$-quark in the unpolarized parton
distribution functions of the proton.
Undoubtedly, the two physics cannot be completely unrelated.

\section{Conclusion}

To conclude, we have given a theoretical prediction for the forward
limit of the isovector, spin flip generalized parton distribution
function $E^{(I = 1)} (x, \xi, t)$ of the nucleon on the basis of
the CQSM. It has been shown that the distribution function
$E_M^{(I = 1)} (x, 0, 0) \equiv H^{(I = 1)} (x, 0, 0) + 
E^{(I = 1)} (x, 0, 0)$ has a sharp peak around $x = 0$ generated by
the vacuum polarization of the Dirac sea quark.
In view of the fact that the function $E_M^{(I = 1)} (x, 0, 0)$
gives the distribution of the nucleon isovector magnetic moment
in the Feynman momentum $x$ space, we interpret this sharp peak
around $x = 0$ as an indication of the importance of the pionic
$q \bar{q}$ excitation with large spatial extension in the transverse
direction. Somewhat unexpectedly, we found that the second
moment of $E_M^{(I = 1)} (x, 0, 0)$ does not reduce to the proton
matrix element of the free quark angular momentum operator, but
receive a peculiar correction term.
The cause of this correction term seems
to be traced back to the nonperturbative formation of isospin
dependent hedgehog mean field. Still, we advocate a viewpoint that
the second moment of $E_M^{(I = 1)} (x, 0, 0)$ gives the isovector
quark angular momentum fraction of an interacting theory,
which leads us to an interpretation of
$\frac{1}{2} x E_M^{(I = 1)} (x, 0, 0)$ as the isovector 
quark (and antiquark) angular momentum distribution.
This then indicates that the $\bar{d}$-quark carries more angular
momentum than the $\bar{u}$-quark inside the proton.

\begin{acknowledgments}
This work is supported by a Grant-in-Aid for Scientific
Research for Ministry of Education, Culture, Sports, Science
and Technology, Japan (No.~C-16540253)
\end{acknowledgments}

\newpage

\end{document}